\documentclass[pre,showpacs,twocolumn]{revtex4-1}
\usepackage{psfrag}
\usepackage{amssymb,amsmath,amsthm,color}
\usepackage{graphicx}
\usepackage[normalem]{ulem}

\newcommand{\beq}{\begin{equation}}
\newcommand{\eeq}{\end{equation}}
\newcommand{\derpar}[2]{\frac{\partial #1}{\partial #2}}
\newcommand{\ve}[1]{{\bf #1}}
\newcommand{\ha}[1]{{\bf \hat #1}}

\begin{document}
\title{Self-assembly of Active Colloidal Molecules with Dynamic Function} 

\author{Rodrigo Soto}
\affiliation{Departamento de F\'{\i}sica, Facultad de Ciencias F\'{\i}sicas y Matem\'aticas Universidad de Chile,
Av. Blanco Encalada 2008, Santiago, Chile}

\author{Ramin Golestanian}
\affiliation{Rudolf Peierls Centre for Theoretical Physics, University of Oxford, Oxford OX1 3NP, UK}

\begin{abstract}
Catalytically active colloids maintain non-equilibrium conditions in which they produce and deplete
chemicals and hence effectively act as sources and sinks of molecules. While individual colloids
that are symmetrically coated do not exhibit any form of dynamical activity, the concentration fields
resulting from their chemical activity decay as $1/r$ and produce gradients that attract or repel other
colloids depending on their surface chemistry and ambient variables. This results in a non-equilibrium
analogue of ionic systems, but with the remarkable novel feature of action-reaction symmetry breaking.
We study solutions of such chemically active colloids in dilute conditions when they join up to form
molecules via generalized ionic bonds, and discuss how we can achieve structures with time dependent
functionality. In particular, we study a molecule that adopts a spontaneous oscillatory pattern
of conformations, and another that exhibits a run-and-tumble dynamics similar to bacteria. Our study
shows that catalytically active colloids could be used for designing self-assembled structures
that posses dynamical functionalities that are determined by their prescribed 3D structures,
a strategy that follows the design principle of proteins.
\end{abstract}

\date{\today}

\maketitle

\section{Introduction}

It is fascinating to learn how Nature exploits sophisticated mechanisms to keep time \cite{Winfree:2001}, since having a system with temporal structure such as spontaneous oscillations in an over-damped inertia-less world is far from trivial. Oscillators are ubiquitous in biology \cite{Vilfan:2005}, and they often play a vital role, such as the example of mitotic spindle oscillations that regulate the key function of chromosome separation during cell division \cite{Howard}. Oscillatory behaviour typically arises through coordination of many stochastic components \cite{Julicher:1997}, and could have remarkable characteristics, such as self-tuning to criticality in the case of hair-bundles \cite{Camalet:2000,Martin:2003,Bruinsma:2013}. Another notable example of temporal structure is the run-and-tumble behaviour in the swimming pattern of microorganisms: the trajectories have relatively long {\em run} segments that are intercalated with burstlike {\em tumble} events when the orientation of the swimming microorganism is completely randomized \cite{berg1}. This behaviour has been extensively studied for many bacteria that are propelled by rotary motors and stiff flagella, but has also been observed for eukaryotic alga that swim via synchronized beating of flexible flagella \cite{ptdgg:2009,BennettGolestanian}. For {\em E. coli}, the coupling between a highly sensitive chemotactic circuitry and the motility mechanism has been unraveled to exquisite detail \cite{Sourjik:2004}.

It will be desirable to make artificial microscopic devices with autonomous dynamic functionality, such as those described above, through self-assembly of nano-scale building blocks, as is the case with the biological examples. Since the turn of the century, there have been numerous manifestations of synthetic devices with mechanical functionality using a number of different approaches, including self-assembled DNA nano-structures \cite{Yurke:2000,Bath:2007}, actuators that are triggered by global oscillatory chemical reactions \cite{Tabata:2002,Kuksenok:2011}, and magnetically actuated artificial micro-swimmers \cite{Dreyfus_nature:2005,Wang:2011} and artificial cilia \cite{Vilfan:2010,Shields:2010,Coq:2011}. A particular class of such active systems \cite{Kapral:2014} that uses long-range phoretic interactions \cite{Paxton:2004,Howse:2007} has been shown to lead to the emergence of collective activities including 
pattern formation and spontaneous oscillations \cite{Ibele2010,Saha:2014}, as well as swarming \cite{Cohen:2014}, in the case of homogenous solutions. In heterogenous mixtures, chemotactic interactions could lead to spontaneous formation (self-assembly) of active colloidal molecules with nonequilibrium activity or {\em function} that is determined by their stable 3D {\em structure}; a notion that follows the design principle of proteins \cite{activemolecules}.

The model for self-assembled active molecules presented in Ref. \cite{activemolecules} was used to demonstrate the formation of low weight colloidal molecules that exhibit different types of static activity, namely translational and rotational self-propulsion. The type of activity depends on the global parameters and the number of colloids that composed each assembly. Here we show that the model allows for the self-assembly of colloidal molecules that adopt time-dependent configurations, increasing the possible states that active matter can manifest. Specifically, we show that a molecule can switch spontaneously between active and passive states, mimicking a behaviour that resembles that of bacteria in their run-and-tumble motion. Also, we show that colloidal molecules can sustain spontaneous oscillations with frequencies that can be tuned by the global parameters, and hence act as microscopic clocks.

\section{The Model}

We consider a suspension of spherical colloidal particles that have a catalyst coating on their surfaces, in a solution of reactants that are catalytically converted into products at the surfaces of the colloids. We choose a mixture of particles with uniform coating, and, for simplicity, use a model in which the catalytic activities of the colloids are simplified into net production or consumption of chemicals with given rates, which we denote as surface activity \cite{gla2005,ruckner2007,gla2007,popescu3,udo}. An isolated colloid with surface activity $\alpha$, which is positive (negative) when the net activity amounts to production (consumption) of chemicals, will produce a stationary spherically symmetric concentration profile $C$ around it given by $C=C_0+\alpha \sigma^2/(4 D r)$, where $\sigma$ is the diameter of the colloid, $D$ is the (nominal) diffusion coefficient of the chemicals, $r$ is the distance to the centre of the colloid, and $C_0$ is a reference concentration at infinity. Since the Brownian diffusivity of the colloid, which we denote as $D_c$, is much smaller than $D$, the solute concentration profile relaxes very quickly to a comoving cloud when a colloidal particle moves. For simplicity, we ignore the effect of advection, which at finite P\'eclet numbers will distort this cloud, because it will affect the self-propulsion velocity only within a pre-factor of order unity \cite{Sharifi:2013}. We also ignore the possibility of spontaneous symmetry breaking at large P\'eclet numbers \cite{lauga} and the anomalous super-diffusion at relatively short time scales \cite{MSD}.

In presence of a concentration gradient a coated particle will move with a velocity equal to $\ve V=-\frac{\mu}{\pi \sigma^2} \int dS\, \nabla_\parallel C$, where $\mu$ is the phoretic mobility, which itself depends on the molecular interactions between the coated surface and the dissolved chemical. As in the case of $\alpha$, this coefficient too can be both positive and negative. Due to these phoretic effects, if two or more colloidal particles are placed close to each other, they will acquire drift velocities that can be regarded as effective nonequilibrium interactions. Remarkably, these interaction are asymmetric as the gradient each particle creates is controlled by its $\alpha$ whereas its response the gradient created by others is controlled by $\mu$; two independent material parameters. More explicitly, the drift velocity of particle 2 due to the activity of particle 1 will be proportional to $\alpha_1\mu_2$ whereas the drift velocity of particle 1 due to the activity of particle 2 will be proportional to $\alpha_2\mu_1$. When the effective interactions between the particles are not symmetric, the system cannot reach an equilibrium state because the condition of detailed balance will not be fulfilled. This can manifest itself in the form of frustration that leads to nonequilibrium fluxes. For example, if $\alpha_1 > 0$ and $\mu_1 > 0$, while $\alpha_2 > 0$ and $\mu_2 < 0$, 1 will be repelled by 2 but 2 will be attracted by 1. In a mixture, we generically have $\alpha_1\mu_2\neq\alpha_1\mu_2$ and, hence, nonequilibrium colloidal activity. For dissimilar colloids, the condition of detailed balanced can only be satisfied by fine-tuning.

In the far-field approximation the drift velocity of particle 2 due to the activity of particle 1 is equal to
\beq
\ve V_2 = -\frac{\mu_2}{\pi \sigma^2} \int dS \, \nabla_\parallel \left( \frac{\alpha_1 \sigma^2}{4 D r}\right)= \frac{\alpha_1\mu_2 \sigma^2}{24 \pi D} \frac{\ve {r}_{12}} {|\ve {r}_{12}|^3}, \label{ABinteraction}
\eeq
where $\ve r_{12}=\ve r_2-\ve r_1$. The effective interaction is similar to unscreened electrostatic interaction between charged colloidal particles in a fluid. Hence, we can regard the two parameters as generalized charges: $\alpha$ is responsible for the production of the field and $\mu$ controls the response to the field. When we are not in the far-field limit, the concentration profile should be calculated by solving the diffusion equation with the appropriate boundary conditions, and the resulting drift velocities will be modified in the near-field region and in general will not be pairwise additive. However, these many-body and proximity effects will only quantitatively change the predictions that emerge from an analysis based on the calculation of the interactions in the far-field limit, where it is pairwise additive, and not affect the qualitative features. In particular, those effects do not alter the main property of the colloidal interactions, namely, the asymmetry of action versus reaction. The far-field approximation of the Coulomb interactions augmented with short range steric repulsion between the particles make the basis of the highly successful and widely used restricted primitive model (RPM) for charged colloids \cite{YanLevin}. Here we adapt this model to our nonequilibrium generalized Coulomb-like system. We also ignore hydrodynamic interactions between colloids as they respect the action-reaction symmetry and appear as sub-dominant terms in the far-field limit. In the near field, the hydrodynamic interactions will modify the attraction or repulsion speeds between colloids due to lubrication effects, changing only quantitatively the predictions made using this simplified model.

We thus perform a Brownian dynamics simulation of active colloidal particles by solving the following stochastic equations
\begin{equation}
\frac{d \ve {r}_i}{d t}=V_0  \sum_{k\neq i} \widetilde{\alpha}_k\widetilde{\mu}_i
\frac{\sigma^2 \ve {r}_{ki}}{|\ve {r}_{ki}|^3} + \sum_{k\neq i}\ve U_{ik}+\boldsymbol{\xi}_i(t), \label{eq.fullmodel}
\end{equation}
where $\boldsymbol{\xi}$ is a random velocity represented by white noise of intensity $2 D_c$ with $D_c$ being
the (passive) diffusion coefficient of the colloidal particles, and $\ve U_{ik}=-\ve U_{ki}$ is the steric repulsion term that prevents particles from overlapping. Using $\alpha_0$ and $\mu_0$ as characteristic values for the surface parameters, we have made the surface activity and mobility parameters dimensionless by defining $\widetilde{\alpha}=\alpha/\alpha_0$ and $\widetilde{\mu}=\mu/\mu_0$. This gives us an overall velocity scale of $V_0=\frac{\alpha_0 \mu_0}{24 \pi D}$ and the dimensionless noise intensity of $\widetilde D_c=D_c/V_0\sigma$, which we regard also as a dimensionless temperature. For simplicity the interactions are computed assuming that the solute molecules could diffuse in three dimensions (with the effective potential decaying as $1/r$) but they colloids are restricted to move in two dimensions. To simulate the excluded volume term, we follow the strategy devised in Ref. \cite{Brownian}. We advance the system in time steps of $\delta t =0.001\sigma/V_0$, and at each time step, compute the drift velocity for each particle by summing all pairwise interactions. The particles are then advanced using a forward Euler scheme that adds the corresponding noise term. At this stage, we identify pairs of overlapping particles and reflect them, in order, by the same distance that they overlap. We repeat the procedure until there are no remaining overlaps.

In what follows we consider a binary mixture, with both species---labelled $A$ and $B$---having the same diameters, but different charges. We explore the case in which $A$ and $B$ particles are mutually attracted (although with different intensities), while identical particles repel each other. The characteristic values $\alpha_0$ and $\mu_0$ are chosen such that for $B$ particles $\widetilde{\alpha}_B=\widetilde{\mu}_B=-1$  and $\widetilde{\mu}_A\geq0$. Without loss of generality, we choose $\widetilde{\alpha}_A\geq 1$ (the opposite case is obtained by exchanging the role of $A$ and $B$ particles).

\section{Oscillatory Instability}

\begin{figure}[b]
\includegraphics[width=.8\columnwidth]{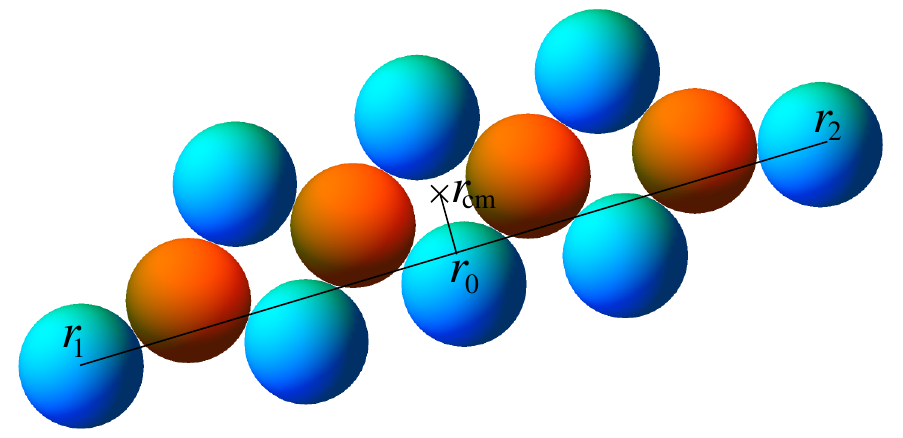}
\caption{Configuration of a molecule with polar and plane-reflection symmetries, where the colloids are free to move with respect to one another subject to the constraint that $A$ and $B$ colloids remain in contact. It has $N_A=4$ and $N_B=8$ colloids and there are 10 degrees of freedom. The oscillations of this molecule are characterized by the parameter $\chi=\big ( [\ve{r}_1-\ve{r}_2]\times[\ve{r}_{cm}-\ve r_0] \big )\cdot \ha{z}$, where $\ve{r}_{cm}$ is the position of the centre of mass, $\ve{r}_{1/2}$ are the positions of the two extreme particles, and $\ve r_0= (\ve{r}_1+\ve{r}_2)/2$, as labelled in the figure.}
\label{fig.oscilaconfig}
\end{figure}

The absence of action-reaction symmetry in the effective interaction between colloids implies that the system does not necessarily evolve toward minima of effective nonequilibrium free energies: it is in principle possible to have residual dynamics in the long time limit, e.g. in the form of limit cycles. The limit cycles will be asymptotic time-dependent stable solutions where the system may be trapped (despite the Brownian noise) until large perturbations occur, possibly in the form of collisions of active molecule with other molecules in the present case.

\begin{figure}[b]
\includegraphics[width=.9\columnwidth]{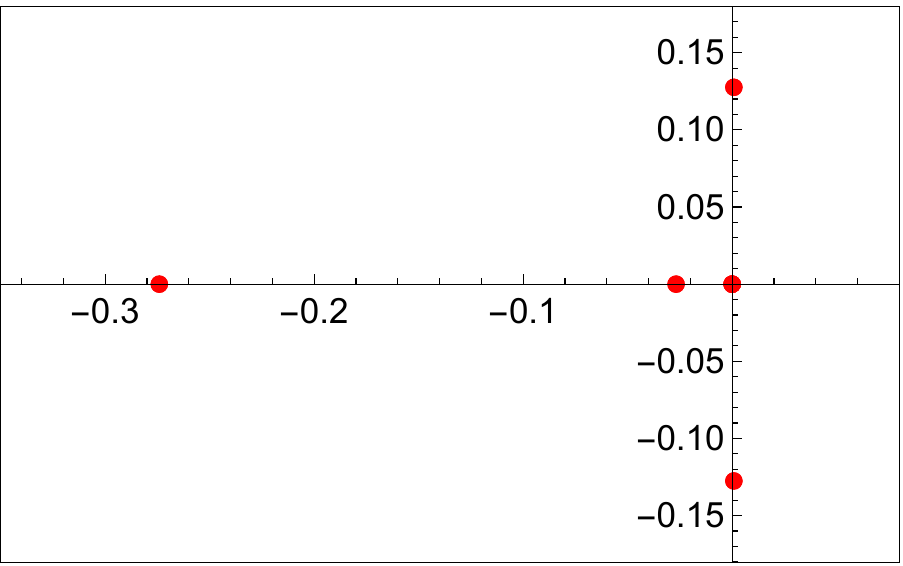}
\caption{Real and imaginary parts of the dimensionless eigenvalues $\lambda$ of the linearized motion of the $A_4B_8$ molecule (shown in Fig. \ref{fig.oscilaconfig}) for $\widetilde\alpha_A=1.5$ and $\widetilde\mu_A=0.7$, where two complex conjugate eigenvalues cross the imaginary axis, with imaginary parts of $\pm 0.127$. Only the region close to the imaginary axis is shown. Not shown are the three remaining eigenvalues, which are real and negative. The null eigenvalue has degeneracy three. Increasing $\widetilde{\alpha}_A$ or decreasing $\widetilde{\mu}_A$ moves the two critical eigenvalues to the positive real region, inducing an instability that is saturated with non-linear terms leading to the development of a limit cycle at the Hopf bifurcation.}
\label{fig.eigenvalues}
\end{figure}

\begin{figure}
\includegraphics[width=0.9\columnwidth]{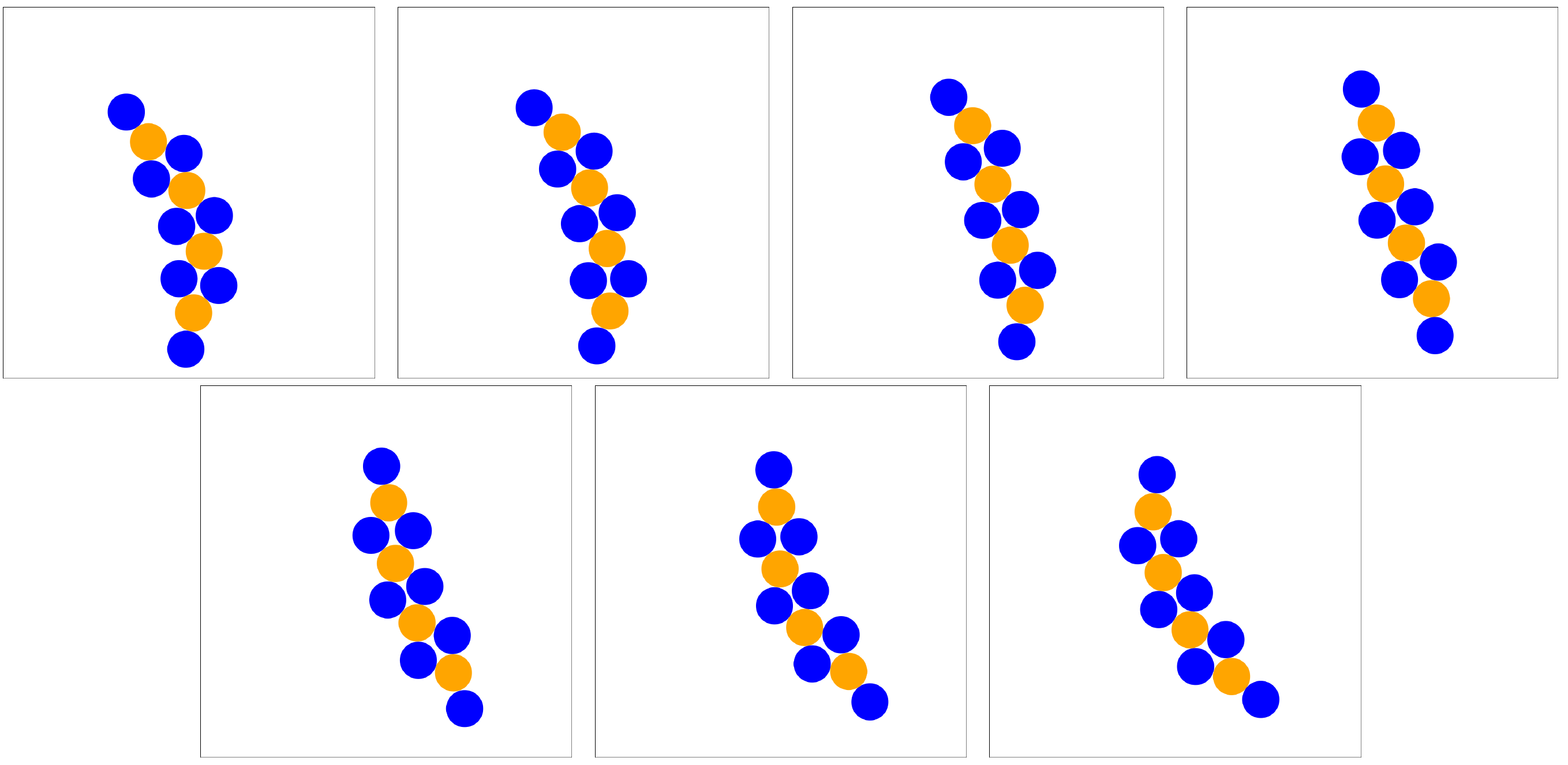}
\caption{Configurations of the molecule that undergoes oscillatory motion at different phases of the cycle, with $\widetilde\alpha_A=1.5$ and $\widetilde\mu_A=0.6$. The full motion is presented in the supplementary material \cite{supmat}.
}
\label{fig.oscilasnaps}
\end{figure}

We are interested in finding low weight molecules that can sustain stable oscillatory motion. To isolate the oscillation from other kinds of activity, we  consider molecules with polar and plane-reflection symmetries to avoid occurrence of translational and rotational self-propulsion. Figure \ref{fig.oscilaconfig} presents a sample configuration of the molecule with the aforementioned symmetry, which we will consider. As $A$ and $B$ colloids mutually attract, they will remain in contact if the noise intensity is small. Therefore, although simulations are performed with full dynamics, we can introduce a further simplification for the purpose of performing the analysis, and consider the case where the particles are restricted to remain in contact and use d'Alembert's principle to derive the dynamics of the remaining coordinates (see Appendix \ref{app.dalembert}). To parameterize the motion of this molecule we chose the following generalized coordinates: the $x$ and $y$ coordinates of the leftmost colloid and all the angles that give the position of the colloids relative to the previous one, going from left to right, always preserving the condition that $A$ and $B$ colloids are in contact. The equations of motion that result after using d'Alembert's principle are first solved to find the equilibrium configuration (regardless of it stability). Then, the equations are linearized around equilibrium, which are analyzed to find relaxation modes with exponential time dependence $e^{V_0\lambda t/\sigma}$. The dimensionless eigenvalues $\lambda$ depend on the values of the charges, and we find that a Hopf bifurcation can take place. Figure \ref{fig.eigenvalues} present the eigenvalues at the bifurcation, where a pair of complex eigenvalues cross the imaginary axis, and acquire positive real parts. There are three null eigenvalues associated to the translational and rotational symmetries. At the bifurcation all other eigenvalues are negative, corresponding to damped motion. Consequently, as it is generically the case, in the neighborhood of the Hopf bifurcation the stable fixed point associated with the straight molecule gives birth to a small stable limit cycle. The limit cycle corresponds to a bending oscillation of the molecule, which, in turn, generates a periodic oscillation of the center of mass in the transverse direction. No rotation is obtained and the net translation in one cycle vanishes. Figure \ref{fig.oscilasnaps} displays the configurations at different phases of the cycle and the full motion is presented in the supplementary material \cite{supmat}. The period of the oscillation is given by the imaginary part of the two critical eigenvalues and the amplitude of the oscillation increases with the distance to the critical point. A movie of the self-assembly process that leads to the formation of the $A_4B_8$ molecule from an initial random dispersion of colloids is presented in the supplementary material \cite{supmat}.

\begin{figure}
\includegraphics[width=.9\columnwidth]{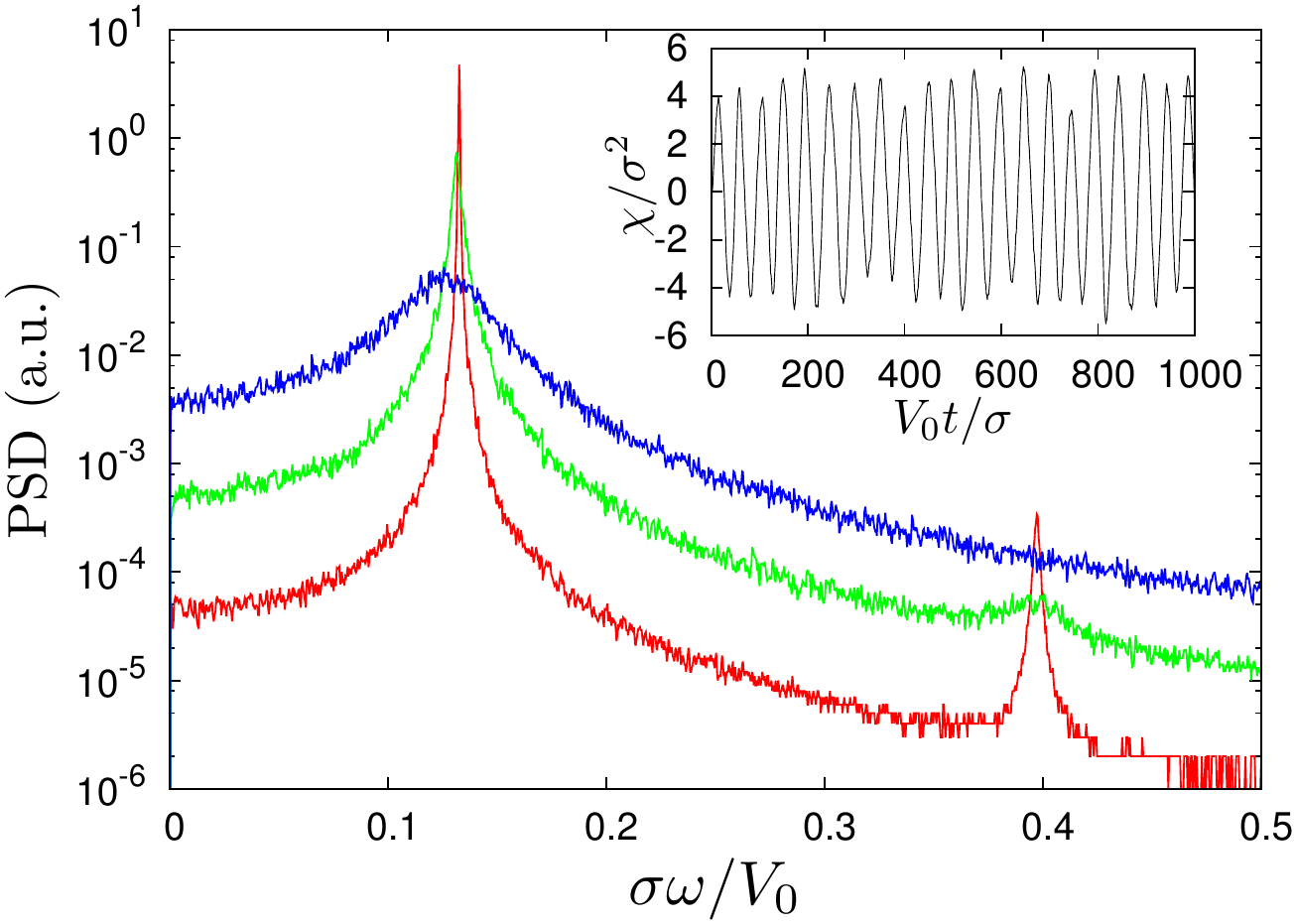}

\includegraphics[width=.9\columnwidth]{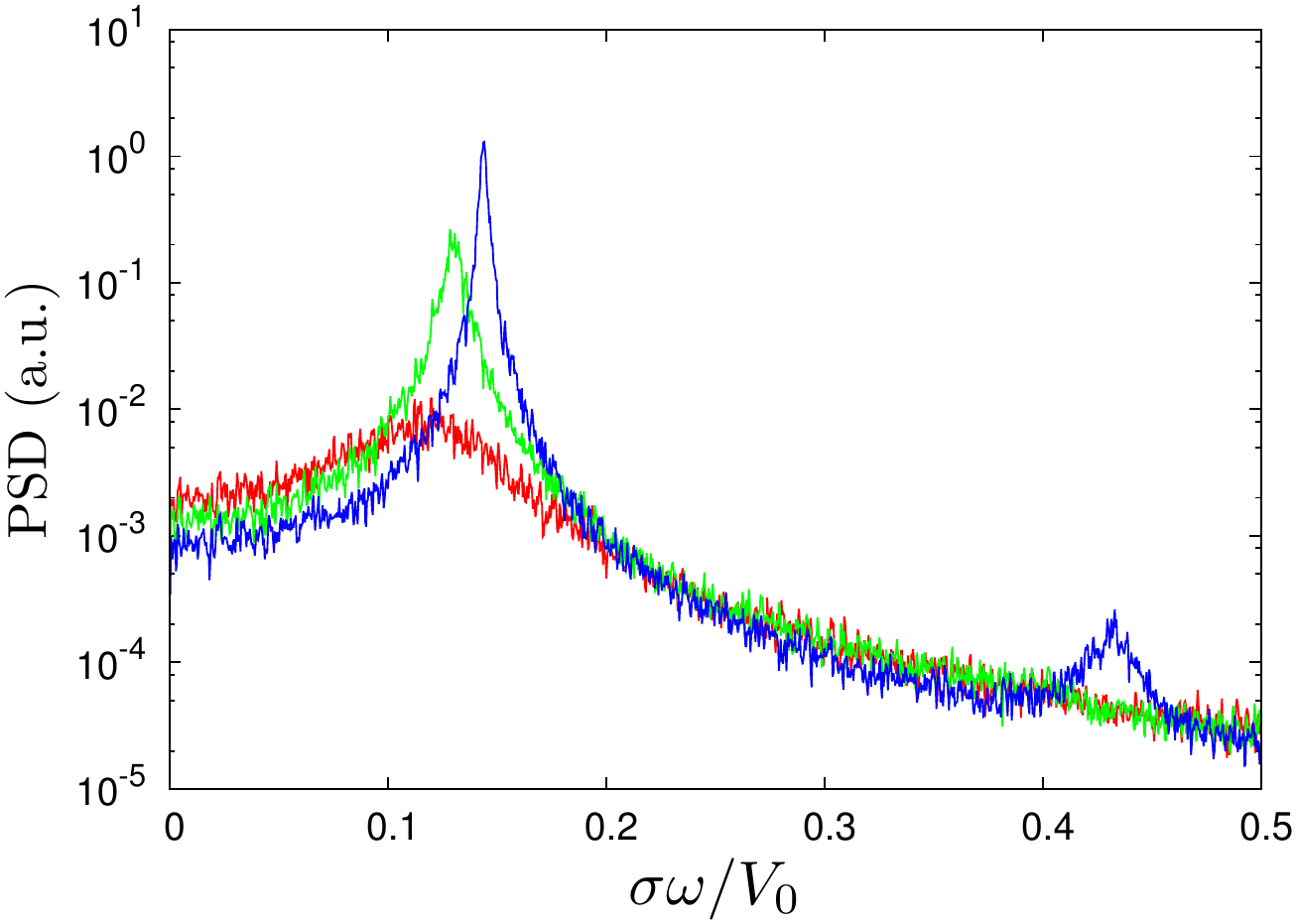}
\caption{Top: Power Spectrum Density (PSD) of $\chi$ in arbitrary units for $\widetilde{\alpha}_A=1.5$ and $\widetilde{\mu}_A=0.6$, and corresponding to $\widetilde{D}_c=5\times10^{-5}$ (red), $\widetilde{D}_c=5\times10^{-4}$ (green), and $\widetilde{D}_c=2.5\times10^{-3}$ (blue). (Inset) Temporal evolution of $\chi$, which measures the instantaneous deformation of the cluster, for $\widetilde{\mu}_A=0.6$ and $\widetilde{D}_c=5\times10^{-4}$.
Bottom: Power Spectrum Density of $\chi$ in arbitrary units for $\widetilde{\alpha}_A=1.5$ and $\widetilde{D}_c=10^{-3}$, and corresponding to $\widetilde{\mu}_A=0.7$ (red), $\widetilde{\mu}_A=0.6$ (green), and $\widetilde{\mu}_A=0.5$ (blue).
}
\label{fig.powespect}
\end{figure}

The quality of the microscopic oscillator can be studied by analyzing the temporal evolution of a simple observable. We define $\chi=\big ( [\ve{r}_1-\ve{r}_2]\times[\ve{r}_{cm}-(\ve{r}_1+\ve{r}_2)/2] \big )\cdot \ha{z} $, where $\ve{r}_{cm}$ is the position of the centre of mass, and $\ve{r}_{1/2}$ are the positions of the two extreme particles (see Fig. \ref{fig.oscilaconfig}). This observable measures the instantaneous deformation of the cluster from the linear configuration, while being rotationally and translationally invariant. Figure \ref{fig.powespect}-top presents the typical time evolution of $\chi$ and its power spectrum for different noise intensities. There is a clear peak at $\omega\approx 0.13 V_0/\sigma$, consistent with the frequency predicted by the eigenvalue analysis (Fig. \ref{fig.eigenvalues}). There is a second peak, which corresponds to the third harmonic (there is no second harmonic by symmetry). As expected, the width of the central peak decreases by decreasing the noise intensity. At $\widetilde{D}_c=5\times10^{-5}$ the frequency is $\omega=(0.1326 \pm 0.0002)V_0/\sigma$, whereas for noise intensities larger than $\widetilde{D}_c=2.5\times10^{-3}$, the fluctuations become large and break the cluster, which no longer oscillates. The oscillations also become sharper by moving away from the transition, as it is shown in Fig. \ref{fig.powespect}-bottom, where the power spectrum peak increases by increasing $\widetilde{\mu}_A$.

The colloidal oscillator $A_4B_8$ is the smallest molecule we found where the first eigenvalues that cross the imaginary axis are a complex conjugate pair. For smaller molecules with the same structure, the first eigenvalue to cross the imaginary axis is an isolated real value, producing a stationary instability instead of an oscillatory one. It is nevertheless possible that other smaller molecules with other structures or with more components may present an oscillatory instability. We also note that at equilibrium ($\widetilde\alpha_A=\widetilde\mu_A$) the matrix associated with the linear dynamics is real and symmetric and therefore all eigenvalues are real, and hence oscillations will not be possible as expected.

\section{Run-and-Tumble Motion}

\begin{figure}
\includegraphics[width=.7\columnwidth]{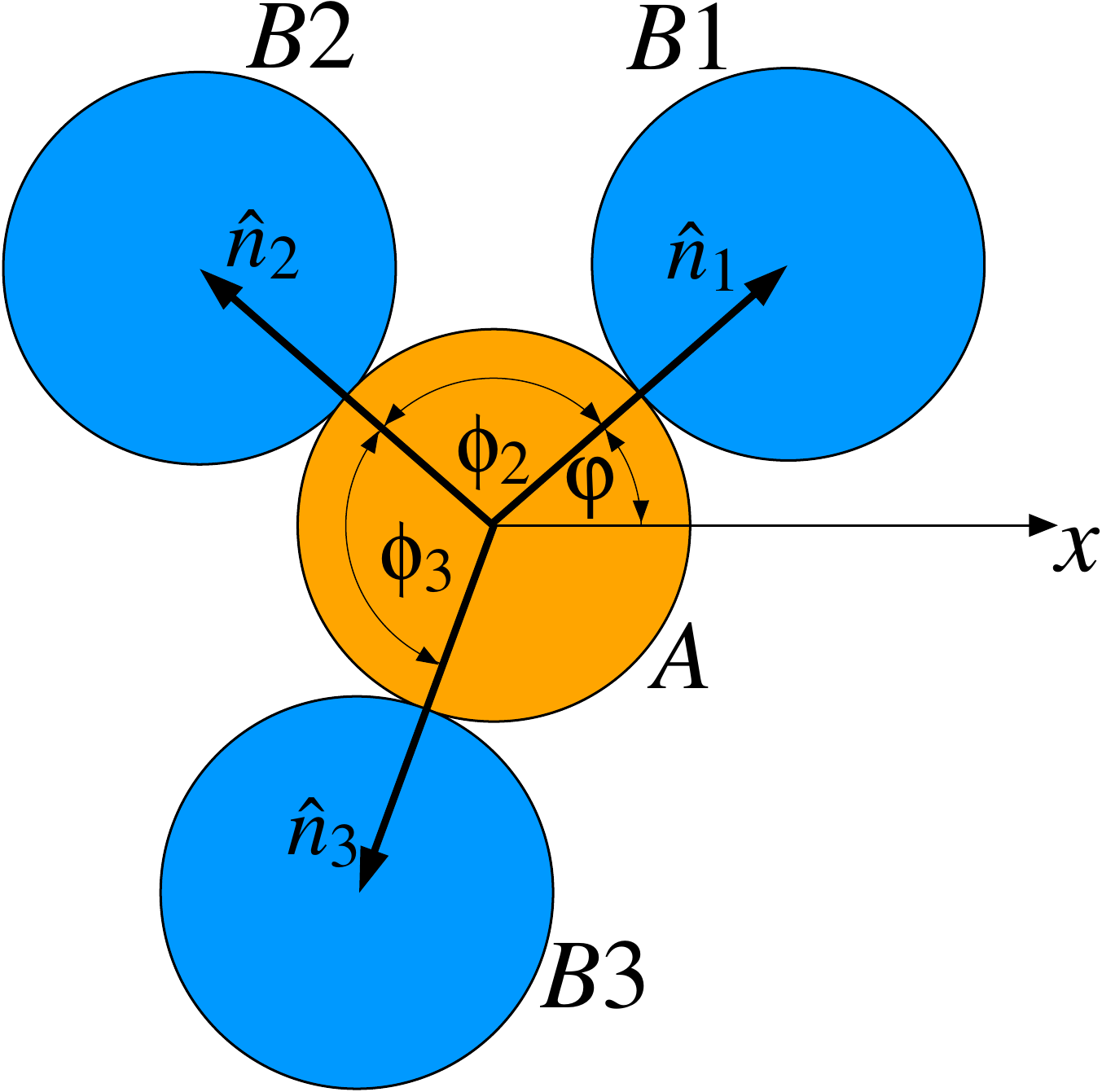}
\caption{Parametrization of the $AB_3$ molecule. The angle between the normal vector $\ha{n}_1$ and the $x$ axis is given by $\varphi$, whereas $\phi_2$ and $\phi_3$ measure the opening angle between consecutive $B$ colloids.}
\label{fig.AB3}
\end{figure}

An alternative way to exhibit dynamic structure is obtained in molecules that could acquire more than one stable configuration, dynamically switching between them. A good example of this behaviour is demonstrated with the $AB_3$ molecule. In a wide range of parameters, this molecule is stabilized with the $A$ colloid located at the centre and the three $B$ colloids maintaining contact with $A$, and otherwise free to move in the angular direction as sketched in Fig. \ref{fig.AB3}. Depending on the parameters, the $B$  colloids can adopt symmetric or asymmetric configurations resulting in isomers than can self-propel or remain Brownian. As in the case of the oscillatory instability, for the purpose of analyzing the possible configurations, their stability, and the transitions between these configurations, we consider the case where the temperature is sufficiently low such that we can restrict the dynamics by imposing that the $B$ colloids always remain in contact with $A$ (i.e. no fluctuations occur in the relative $A-B$ distances). Under this approximation, the equations of motion for each colloid (without the noise terms) are
\begin{align*}
\ve V_A &= (V_0 \widetilde{\mu}_A-U_1) \ha {n}_1 +  (V_0 \widetilde{\mu}_A-U_2) \ha{n}_2 +  (V_0 \widetilde{\mu}_A-U_3) \ha{n}_3 \\
\ve V_{B1} &= -(V_0 \widetilde{\alpha}_A-U_1)\ha{n}_1
+ V_0 \frac{\ha{n}_1-\ha{n}_2}{|\ha{n}_1-\ha{n}_2|^3}
+ V_0 \frac{\ha{n}_1-\ha{n}_3}{|\ha{n}_1-\ha{n}_3|^3}\\
\ve V_{B2} &= -(V_0 \widetilde{\alpha}_A-U_2)\ha{n}_2
+ V_0 \frac{\ha{n}_2-\ha{n}_3}{|\ha{n}_2-\ha{n}_3|^3}
+ V_0 \frac{\ha{n}_2-\ha{n}_1}{|\ha{n}_2-\ha{n}_1|^3}\\
\ve V_{B3} &= -(V_0 \widetilde{\alpha}_A-U_3)\ha{n}_3
+ V_0 \frac{\ha{n}_3-\ha{n}_1}{|\ha{n}_3-\ha{n}_1|^3}
+ V_0 \frac{\ha{n}_3-\ha{n}_2}{|\ha{n}_3-\ha{n}_2|^3},
\end{align*}
where the normal vectors $\ha n_i$ are defined in Fig. \ref{fig.AB3} and $U_i$ are the magnitude of the constraint velocities, which will be eliminated using d'Alembert's principle.

The motion of the centre of mass of the molecule can be readily obtained as
\beq
\ve V = \frac{1}{4}(\ve V_A+\ve V_{B1}+\ve V_{B2}+\ve V_{B3})=(\widetilde{\mu}_A-\widetilde{\alpha}_A)(\ha{n}_1+\ha{n}_2+\ha{n}_3),
\eeq
where it transpires that the molecule propels only under non-equilibrium conditions ($\widetilde\alpha_A\neq\widetilde\mu_A$) and when the conformation is not symmetric ($\ha{n}_1+\ha{n}_2+\ha{n}_3\neq 0$). We now study under which conditions the configuration is not symmetric.

\begin{figure}
\begin{center}
\includegraphics[width=.45\columnwidth]{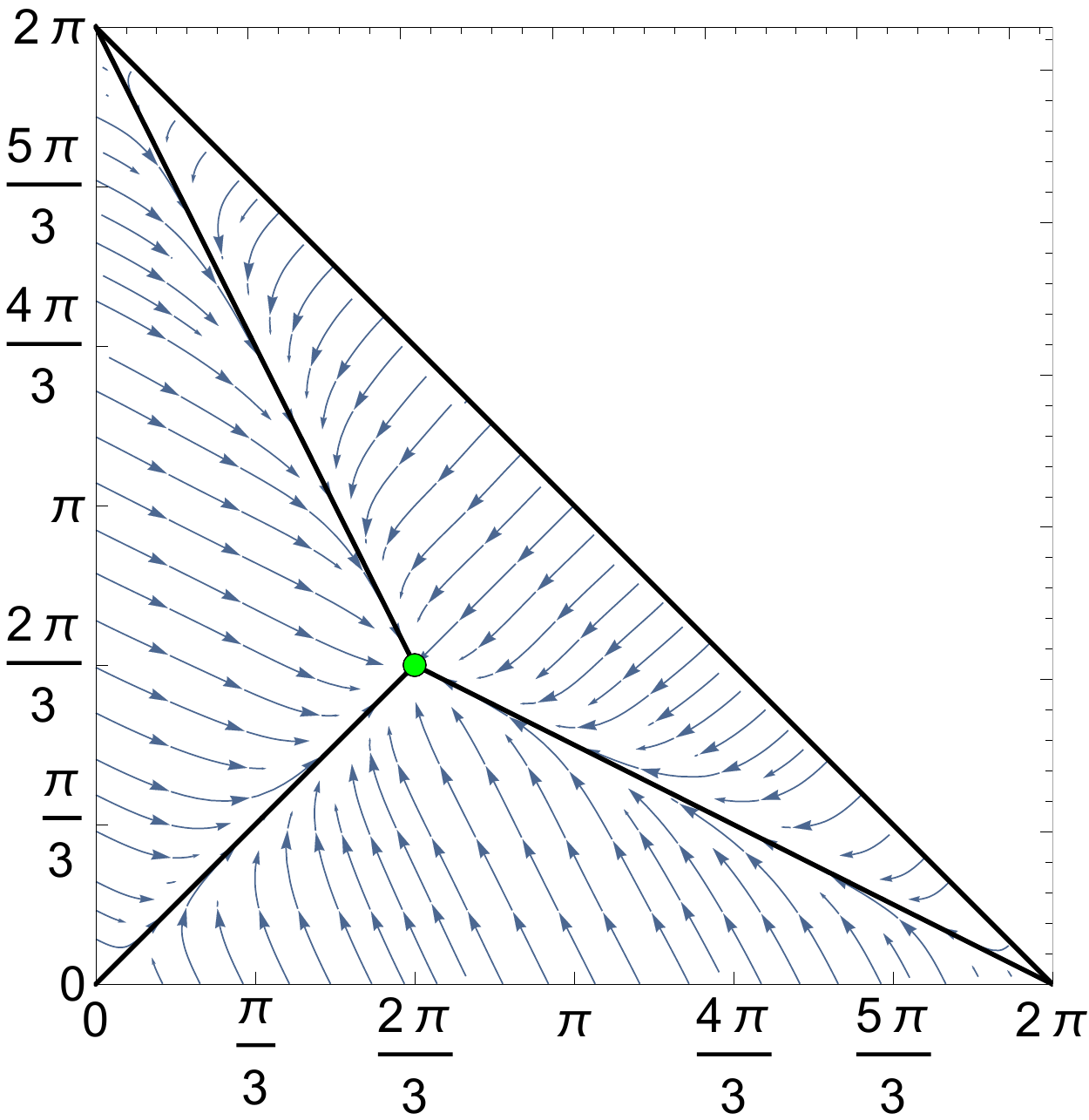}\,
\includegraphics[width=.45\columnwidth]{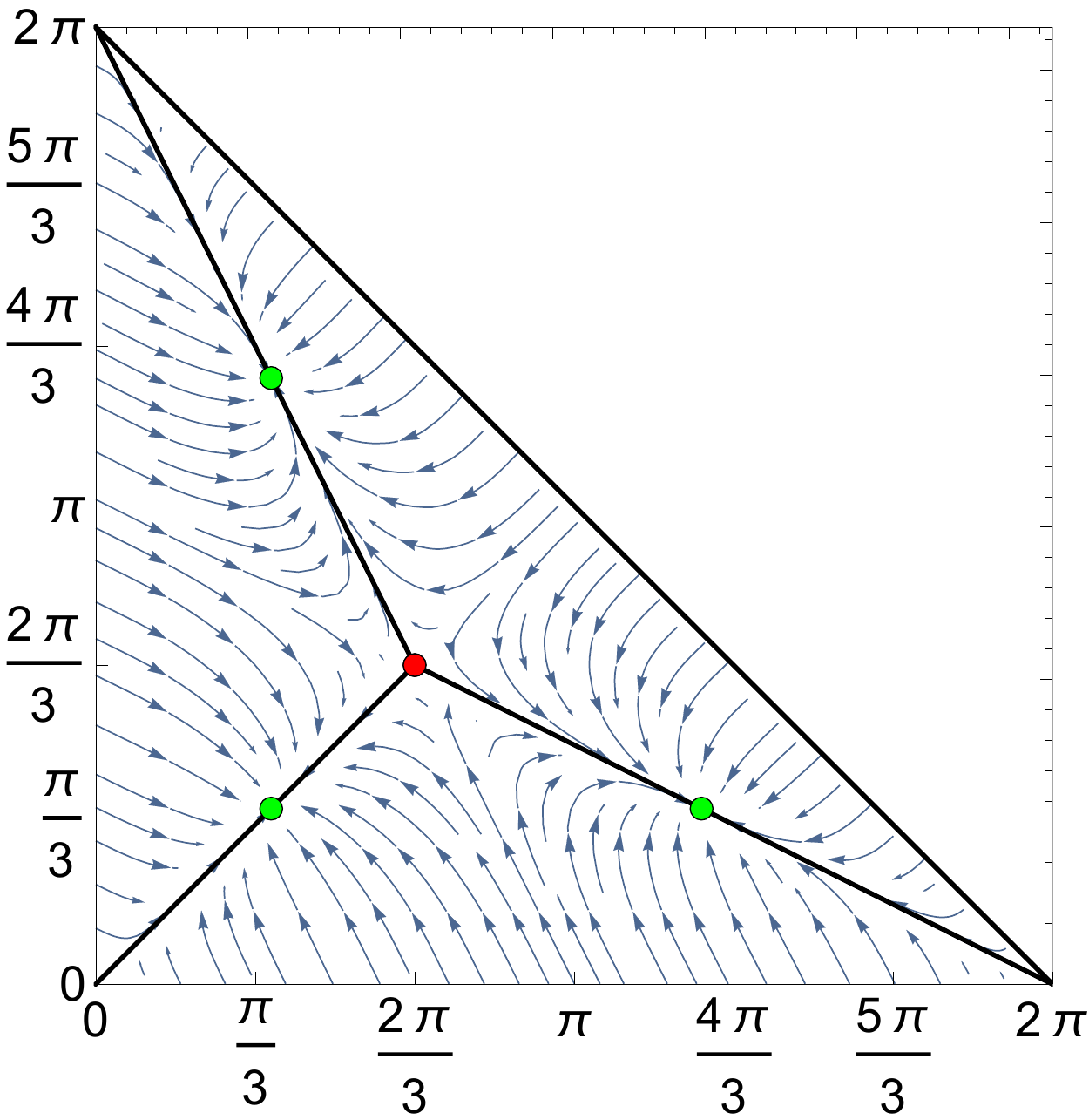}
\includegraphics[width=.9\columnwidth]{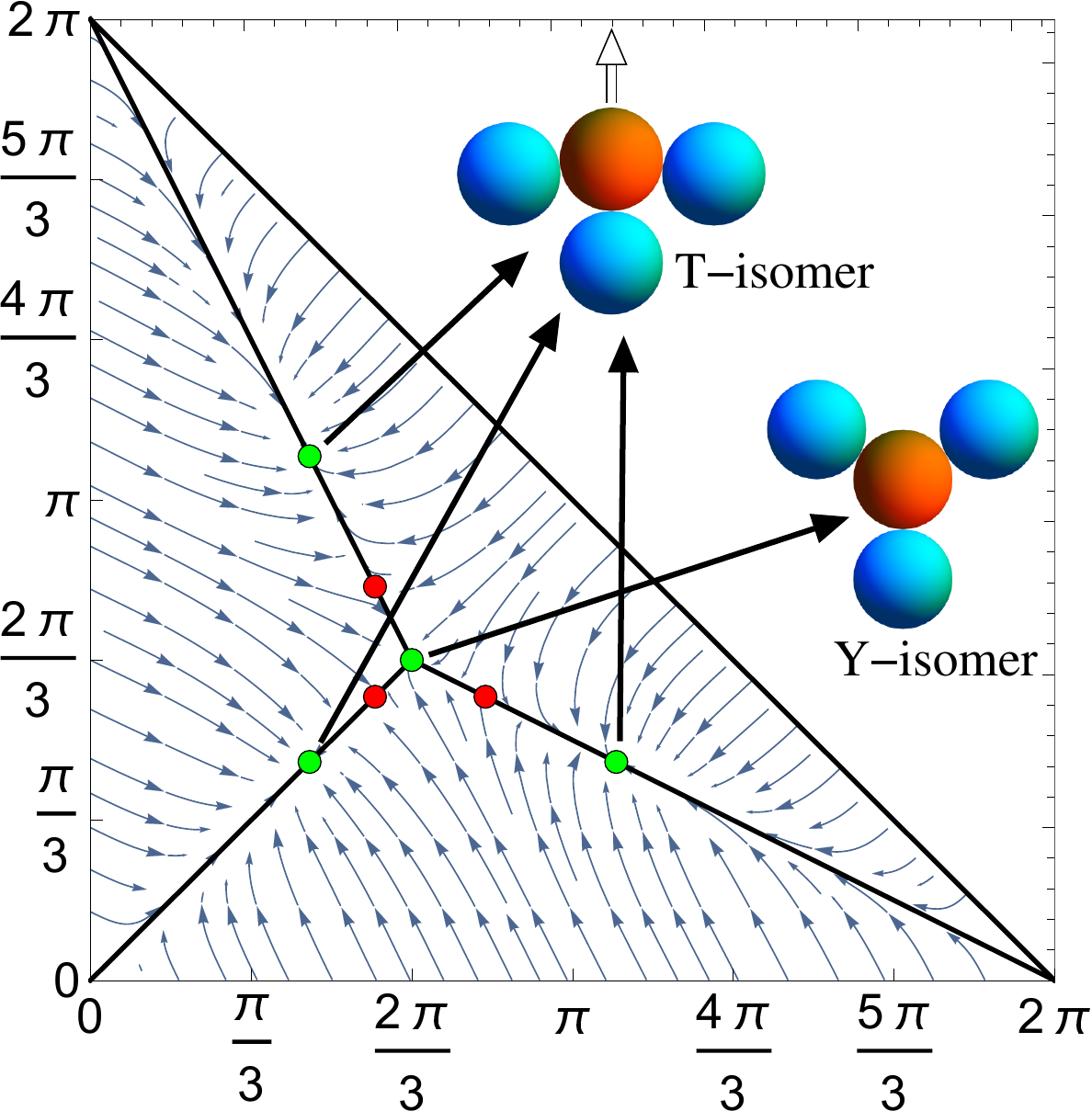}
\end{center}
\caption{Phase portrait in the $\phi_2-\phi_3$ space in the physical domain ($\phi_2+\phi_3\leq2\pi$). The phoretic charges are $\widetilde\mu_A=1$ (top left), $\widetilde\mu_A=0$ (top right), and $\widetilde\mu_A=1/2$ (bottom), while $\widetilde \alpha_A=2$ in the three cases. Stable fixed points are shown in green and saddle points in red. The diagonal solid lines indicate configurations with two equal angles. The stable configurations that correspond to the Y- and T-isomers are indicated by arrows.}
\label{fig.phaseportrait}
\end{figure}

To derive the motion of the molecule, we apply d'Alembert's principle (see Appendix \ref{app.dalembert}), considering the following generalized coordinates for the relevant degrees of freedom: the $x$ and $y$ coordinates of the central $A$ colloid and the three angles $\varphi$, $\phi_2$, and $\phi_3$ defined in Fig. \ref{fig.AB3}. The angle $\varphi$ is cyclic, due to the rotational invariance, implying the conservation law $\dot\varphi_1+\dot\varphi_2+\dot\varphi_3+\dot{y}(\cos\varphi_1+\cos\varphi_2+\cos\varphi_3)-\dot{y}(\sin\varphi_1+\sin\varphi_2+\sin\varphi_3)=0$, where $\varphi_1=\varphi$, $\varphi_2=\varphi+\phi_2$ and $\varphi_3=\varphi+\phi_2+\phi_3$ give the orientation of the three $B$ colloids with respect to the $x$ axis. The remaining equations of motion can be reduced to two coupled equations for $\phi_2$ and $\phi_3$, the bond angles of the molecule: $\dot\phi_2 = \omega_2(\phi_2,\phi_3)$ and $\dot\phi_3 = \omega_3(\phi_2,\phi_3)$. The expressions for $\omega_2$ and $\omega_3$ are straightforward to obtain but are quite involved and we do not show them explicitly. Figure \ref{fig.phaseportrait} presents the phase portrait in the $\phi_2-\phi_3$ space where different fixed points appear depending on the values of the phoretic charge $\widetilde\mu_A$, while keeping $\widetilde\alpha_A$ fixed (a similar picture is obtained by varying $\widetilde\alpha_A$ and fixing $\widetilde\mu_A$). In Fig. \ref{fig.phaseportrait}-top-left, there is only one stable fixed point corresponding to the symmetric {\bf Y-isomer} ($\phi_2=\phi_3=2\pi/3$), which does not self-propel. In Fig. \ref{fig.phaseportrait}-top-right, a second stable equilibrium with triple degeneration appears, corresponding to the asymmetric {\bf T-isomer}, which self-propels in the direction indicated by the double arrow in Fig.\ \ref{fig.phaseportrait}-bottom. Finally, Fig. \ref{fig.phaseportrait}-bottom presents the case of bistability between the two isomers for an intermediate value of $\widetilde\mu_A$. The bifurcation between these phases occurs via the collision of the stable fixed points with the saddle points.

To analyze the transitions between these states in the presence of noise we make a further simplification by noting that both the stable fixed points and the saddle points lie on the lines where the molecule is partially symmetric, with two bond angles being equal. Therefore, we consider the reduced dynamics of a single variable $\phi=\phi_2=\phi_3$, which is governed by the following equation
\begin{align}
\dot{\phi}=&\omega(\phi) +\xi= -(\widetilde\alpha_A-\widetilde\mu_A) \frac{(2\cos\phi+1)\sin\phi}{3+\cos 2\phi}
 \nonumber\\
&+ \frac{\cos\phi/\sin^2\phi + \sqrt{2}\sin\phi/(1-\cos\phi)^{3/2}}{(3+\cos2\phi)} +\xi,
\end{align}
where $\xi$ is the resulting white noise after applying d'Alembert's method, with intensity $2D_c/(2-\sin\phi^2)\sigma^2$. The corresponding Fokker-Planck equation (using Stratonovic convention \cite{Risken}) is
\beq
\derpar{P}{t} = \derpar{}{\phi} \left[-\omega(\phi)P + \frac{D_c/\sigma^2}{\sqrt{2-\sin\phi^2}} \derpar{}{\phi}\left( \frac{P}{\sqrt{2-\sin\phi^2}} \right) \right] \label{FP}.
\eeq

\begin{figure}
\includegraphics[width=.9\columnwidth]{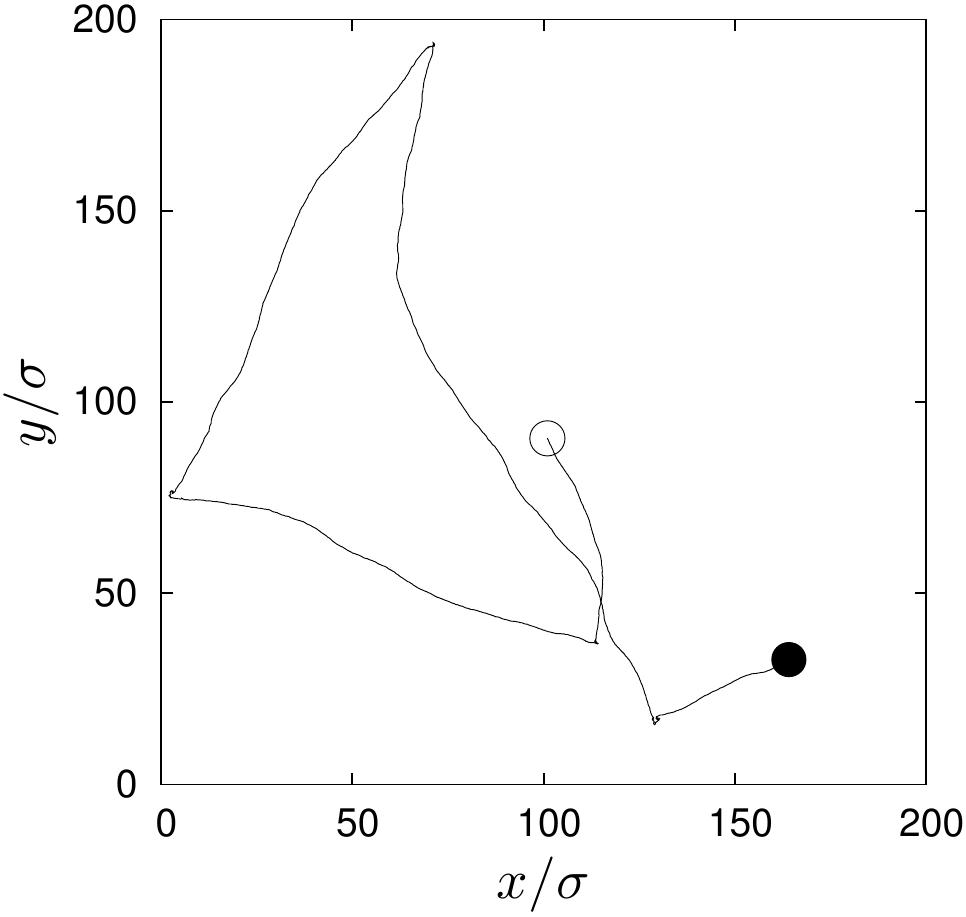}\,
\includegraphics[width=.9\columnwidth]{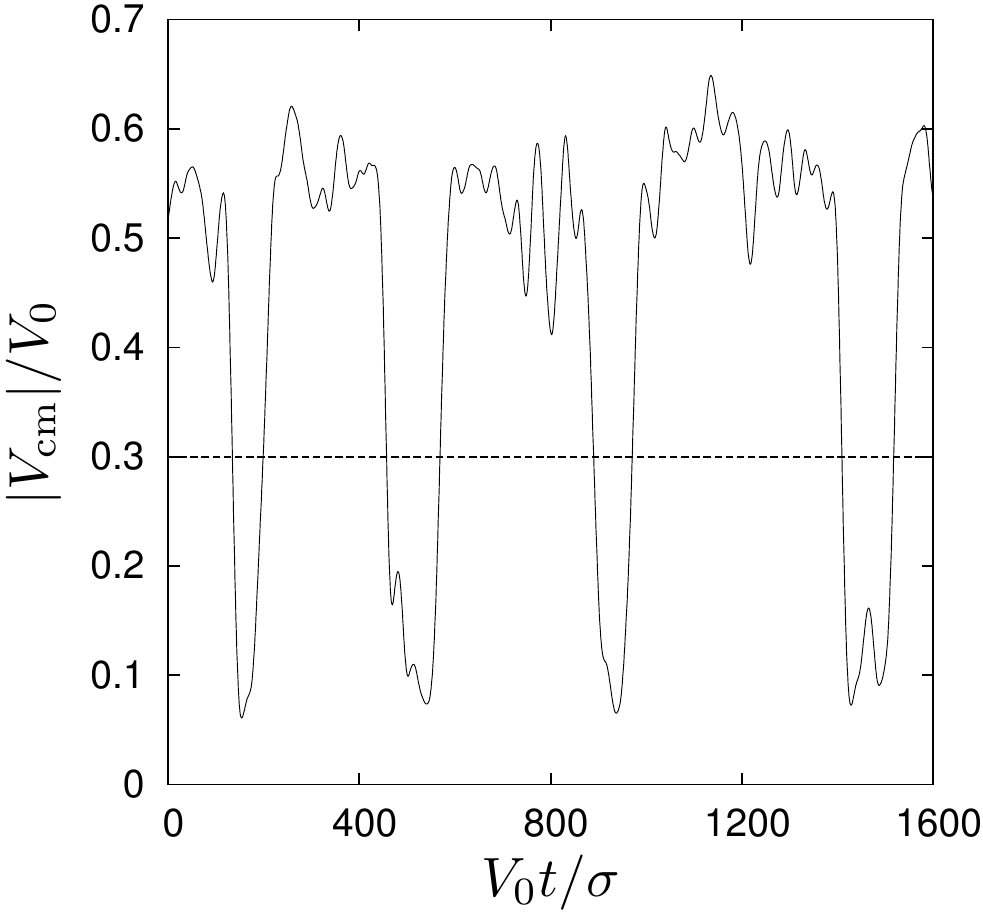}
\caption{Top: Trajectory of the centre of mass showing the run-and-tumble motion of an $AB_3$ molecule for $\widetilde{\alpha}_A=2$, $\widetilde{\mu}_A=0.48$, and $\widetilde{D}_c=2.5\times10^{-3}$. The open and closed circles indicate the beginning and the end of the trajectory, respectively. Bottom: Coarse grained speed of the centre of mass for the trajectory in the top panel. The dashed line indicates the threshold used to discriminate the run and the tumble phases. Five runs and four tumbles are identified as seen in the top panel.
}
\label{fig.RT}
\end{figure}

\begin{figure}
\includegraphics[width=.9\columnwidth]{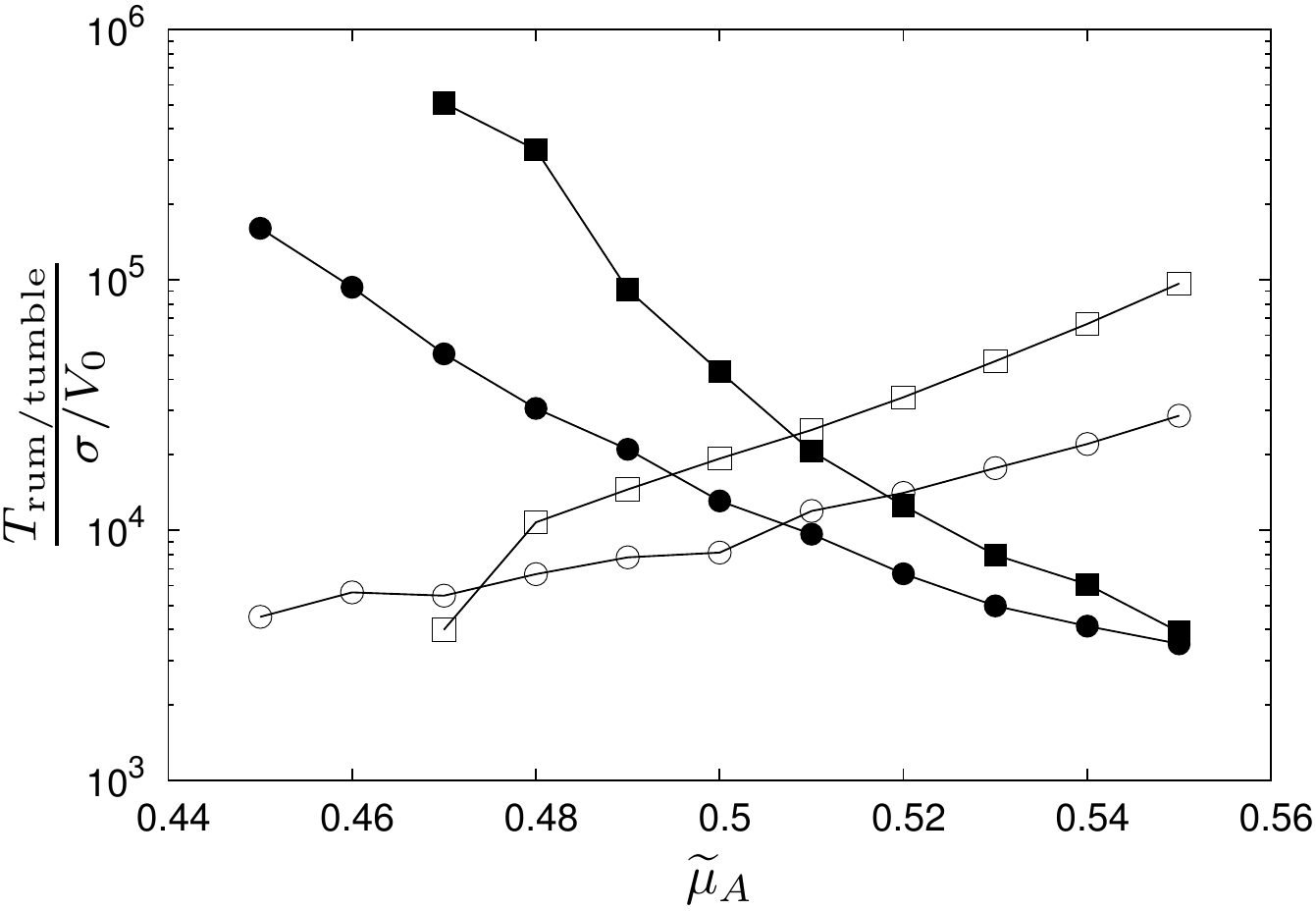}
\caption{Average duration of the run (solid symbols) and tumble (open symbols) phases as a function of $\widetilde{\mu}_A$ for fixed $\widetilde{\alpha}_A=2$. The noise intensities are $\widetilde{D}_c=0.003$ (squares) and $\widetilde{D}_c=0.005$ (circles).
}
\label{fig.RTTimes}
\end{figure}

The resulting stationary distributions that is obtained by solving equation (\ref{FP}) will depend on $\omega$ the $D_c$. However, in the low noise limit, the distributions are peaked around the stable points, which we can obtain from $\omega$ only. We find that in the range $0\leq\widetilde\mu_A<0.0754$ there is only one stable point at a value of $\phi$ which is $<2\pi/3$, and corresponds to the T-isomer. In the range $0.0754\leq\widetilde\mu_A<0.5962$ two stable equilibria exist, corresponding to the coexistence of the T- and Y-isomers. Finally, in the range  $0.5962\leq \widetilde\mu_A$ the only stable point is $\phi=2\pi/3$, namely the Y-isomer.

In the bistable region, the molecule can exist in the two isomeric configurations, making transitions between them due to noise. The dynamics is similar to the run-and-tumble motion performed by swimming bacteria as there is an alternation between run periods performed by the T-isomer that are followed by near pauses during which the molecule acquires the Y-isomer and undergoes rotational diffusion only, leading to the start of a new run phase with a random orientation. Here, the new orientation is decided by two complementary processes. First, in the Y-isomer configuration, there is the rotational diffusion of the cyclic variable $\varphi$. Second, when the molecule transits to the a new T-isomer, it can adopt any of the three degenerate configurations. When the rotational diffusion component is small, tumbling occurs to three possible discrete new orientations, which is in contrast to bacteria that tumble with a continuous distribution of angles \cite{berg1}. A sample trajectory of the centre of mass is shown in Fig. \ref{fig.RT}-top where the run-and-tumble motion is evident, and a video of this motion is presented in the supplementary material \cite{supmat}. The two phases can be discriminated by looking at the coarse grained speed, which adopts values in the vicinity of $0.6 V_0$ during the run phase while it is close to zero in the tumble phase (Fig. \ref{fig.RT}-bottom).

In the bistable region, we have calculated the duration of these phases using the Brownian dynamics simulation by imposing a threshold value on the coarse grained speed (as shown in Fig.~\ref{fig.RT}-bottom). The average times are presented in Fig. \ref{fig.RTTimes}. The residence times in the run and tumble phases exhibit an exponential dependence on the value of $\widetilde\mu_A$, while they increase by reducing the noise intensity. There is a crossover at $\widetilde\mu_A\approx0.51$ where both phases have similar durations. These behaviours are consistent with what we expect from Kramer's first passage time theory \cite{note}.

\section{Concluding Remarks}

We have shown that active colloidal molecules can develop rich dynamical structures, e.g. by having multiple stable fixed points or limit cycles in the space of their conformations. A key ingredient for obtaining these dynamical states is that the colloidal molecules are not rigidly assembled; since they are self-assembled, they are flexible and have internal degrees of freedom that can be active in the same manner as the global translation or rotation modes.

While we focussed on the simplest examples that exhibit temporal structure, we expect more complex temporal patterns of behaviour to appear for larger and more complex molecules. An exemplar movie is presented in the supplementary material where random self-assembly leads to the formation of a sufficiently large oscillator that can break time reversal symmetry and self-propel as a whole while beating its oscillatory tail, exhibiting a pattern of motion that is reminiscent of swimming spermatozoa \cite{supmat}. Further work is needed to study the full range of possible structures that can be achieved via nonequilibrium self-assembly of catalytically active colloids.

The simple model used for the active colloids serves as a proof of concept for self-assembly of molecules with dynamic function, namely micro-oscillators and run-and-tumble motion. The model has several simplifications, notably it neglects hydrodynamic interactions and considers only the far-field phoretic interactions. These and other effects such as specific details of the catalytic reaction kinetics should be included in order to have a realistic description of the system that allows us to make quantitative predictions. Therefore, in practice, it is not guaranteed whether the particular examples presented in this manuscript will perform their time dependent function when all these effects are included. However, these details will only make quantitative changes, which means that similar configurations will exhibit the reported functionality, because those effects do not alter the main underlying mechanism for dynamic function, which is the action-reaction symmetry breaking that is present in the phoretic interactions.


\appendix
\section{D'Alembert's principle in overdamped dynamics} \label{app.dalembert}
Consider an ensemble of particles with overdamped dynamics
\beq
\dot{\ve r}_i = \ve{F}^T_i ; \quad i=1,2,\ldots,N,
\eeq
where $\ve F^T_i=\ve F_i+\ve F^c_i$ is the total force acting on each particle and $\ve F^c_i$ are constraint forces, where the mobilities have been set to one for simplicity. In the case of colloidal particles, these constraint forces keep the relative distances of the particles in contact fixed.

Generalized coordinates $q_k$, $k=1,2,\ldots,n \leq N$, are defined satisfying the constraints such that the particle positions can be uniquely calculated in terms of them; i.e. $\ve r_i =\ve r_i(q_1,\ldots,q_n)$.  With these generalized coordinates virtual displacements $\delta \ve r_i$ can be constructed such that they satisfy, by construction, $\sum_i \ve F^c_i\cdot\delta\ve r_i=0$. Therefore,
\beq
\sum_i \left (\ve F_i-\dot{\ve{r}}_i\right)\cdot\delta\ve r_i=0.
\eeq
Using $\delta\ve r_i=\sum_k\derpar{\ve r_i}{q_k}\delta q_k$, the previous expression can be written as
\beq
\sum_k \left(Q_k - \derpar{}{\dot{q}_k}T \right) \delta q_k=0,
\eeq
where
\beq
Q_k = \sum_i \ve F_i\cdot\derpar{\ve r_i}{q_k}
\eeq
are the generalized forces, which do not depend on the constraint forces,
and
\beq
T = \sum_i \frac{1}{2}\dot{\ve{r}}^2_i
\eeq
plays a role analogous to the kinetic energy.

As the virtual displacements in the generalized coordinates are independent, the resulting equations of motion in the generalized coordinates are
\beq
\derpar{}{\dot{q}_k}T = Q_k ;\quad k=1,2,\ldots,n.
\eeq

\acknowledgments

This research is supported by Fondecyt Grant No. 1100100 and Anillo grant ACT 127 (R.S.),
and Human Frontier Science Program (HFSP) grant RGP0061/2013 (R.G.).


\begin{thebibliography}{99}
\bibitem{note}
To perform a quantitative comparison, we need to solve the Fokker-Planck equation. For the transition from the run to the tumble phase, the analysis can be done using Eq.\ (\ref{FP}) as the transition path is along the imposed one-dimensional trajectory in phase space. For the opposite case, i.e. going from the tumble to a run phase, the analysis must be done with the full equation for $\phi_2$ and $\phi_3$. We relegate this calculation to a future publication.
\end{thebibliography}
\end{document}